\documentclass[lettersize,journal]{IEEEtran}
\usepackage{amsmath,amsfonts}
\usepackage{algorithmic}
\usepackage{algorithm}
\usepackage{array}
\usepackage[caption=false,font=normalsize,labelfont=sf,textfont=sf]{subfig}
\usepackage{textcomp}
\usepackage{stfloats}
\usepackage{url}
\usepackage{verbatim}
\usepackage{graphicx}
\usepackage{cite}
\usepackage{xcolor}
\usepackage[table]{xcolor}
\usepackage[utf8]{inputenc} 
\usepackage[T1]{fontenc}    
\usepackage{hyperref}       
\usepackage{arydshln}     
\usepackage{booktabs}       
\usepackage{nicefrac}       
\usepackage{microtype}      
\usepackage{multirow}
\usepackage{makecell}
\usepackage[numbers]{natbib}
\usepackage{siunitx}
\usepackage{pifont}
\newcommand{\xmark}{\ding{55}}%

\begin{document}

\title{BioME: A Resource-Efficient Bioacoustic \\ Foundational Model for IoT Applications}

\author{
    Heitor R. Guimar\~aes, 
    Abhishek Tiwari, 
    Mahsa Abdollahi, 
    Anderson R. Avila,
    Tiago H. Falk 
\thanks{Received XX February 2026; revised XX March 2026; accepted XX April 2026. Date of publication XX May 2026; date of current version XX February 2026. (Corresponding author: Heitor R. Guimarães; e-mail: heitor.guimaraes@inrs.ca)}

\thanks{The authors are with the Institut national de la recherche scientifique (INRS - EMT), Montréal, QC, Canada.}%

\thanks{Digital Object Identifier XYZ/XX.XXXX.XXXX}
}

\markboth{Journal of \LaTeX\ Class Files,~Vol.~14, No.~8, August~2021}%
{Shell \MakeLowercase{\textit{et al.}}: A Sample Article Using IEEEtran.cls for IEEE Journals}

\IEEEpubid{0000--0000/00\$00.00~\copyright~2021 IEEE}

\maketitle

\begin{abstract}
Passive acoustic monitoring has become a key strategy in biodiversity assessment, conservation, and behavioral ecology, especially as Internet-of-Things (IoT) devices enable continuous in situ audio collection at scale. While recent self-supervised learning (SSL)-based audio encoders, such as BEATs and AVES, have shown strong performance in bioacoustic tasks, their computational cost and limited robustness to unseen environments hinder deployment on resource-constrained platforms. In this work, we introduce BioME, a resource-efficient audio encoder designed for bioacoustic applications. BioME is trained via layer-to-layer distillation from a high-capacity teacher model, enabling strong representational transfer while reducing the parameter count by 75\%. To further improve ecological generalization, the model is pretrained on multi-domain data spanning speech, environmental sounds, and animal vocalizations. A key contribution is the integration of modulation-aware acoustic features via FiLM conditioning, injecting a DSP-inspired inductive bias that enhances feature disentanglement in low-capacity regimes. Across multiple bioacoustic tasks, BioME matches or surpasses the performance of larger models, including its teacher, while being suitable for resource-constrained IoT deployments. For reproducibility, code and pretrained checkpoints are publicly available\footnote{\mbox{\url{https://huggingface.co/collections/Hguimaraes/biome}}}.
\end{abstract}


\section{Introduction}
\label{sec:introduction}

\IEEEPARstart{B}{iodiversity} is the foundation of ecological stability and planetary health, yet it is declining at an unprecedented rate~\cite{butchart2010global}. Accelerated habitat loss, climate change, and other anthropogenically driven pressures have pushed numerous species toward endangerment or extinction, disrupting ecosystem functioning and threatening the socio-economic systems that depend on it~\cite{cardinale2012biodiversity}. Monitoring, preserving, and understanding wildlife and its behavior are therefore not only a matter of ecological ethics but also a global imperative for maintaining food security and long-term human well-being~\cite{ullah2025integrating}.

Leveraging non-invasive sensing, passive acoustic monitoring (PAM) has emerged as a powerful approach in bioacoustics for studying animal-produced sounds~\cite{teixeira2019bioacoustic}, ranging from vocalizations to movement-related signals. Audio carries rich ecological information~\cite{teixeira2024effective} and provides several advantages over other sensing modalities, such as vision. Unlike camera traps, microphones do not require a direct line of sight, are less affected by lighting or occlusion, and can capture multiple individuals simultaneously across long distances. With the advent of the Internet of Things (IoT), autonomous audio recorders have become increasingly affordable, energy-efficient, and deployable at scale, enabling long-term in situ monitoring in remote environments. However, these deployments generate massive volumes of continuous audio data, making manual inspection infeasible and highlighting the need for automatic signal processing and machine learning methods to extract meaningful ecological knowledge.

Neural network-based approaches have set state-of-the-art results in audio classification and detection by learning expressive representations directly from high-dimensional signals. Self-supervised learning (SSL) enables models to exploit the intrinsic structure of data through pretext tasks that do not require human labels. The resulting encoder maps raw audio into a compact latent space where meaningful features are disentangled, facilitating downstream fine-tuning with limited annotations and improving robustness to environmental variability and other confounding acoustic conditions.
\IEEEpubidadjcol

Foundational audio encoders, such as BEATs~\cite{pmlr-v202-chen23ag} and AVES~\cite{hagiwara2023aves}, have recently demonstrated strong performance on different bioacoustic tasks, including bird species classification~\cite{10445889} and whale call detection~\cite{10720021}. Despite their effectiveness, these models are computationally expensive, limiting their deployment on resource-constrained edge devices where near-real-time inference is required. Their size and inference cost may also discourage researchers with limited access to high-end hardware, restricting broader adoption. 
While naively scaling down these models is possible, doing so often results in substantial performance degradation. We argue that maintaining high accuracy under low-capacity constraints requires designing encoders that learn more meaningful and domain-relevant representations, rather than simply reducing model size.

In this work, we introduce BioME (\textbf{Bio}acoustic \textbf{M}odulation-aware \textbf{E}ncoder), a resource-efficient self-supervised audio encoder for bioacoustic tasks. BioME is trained via layer-wise knowledge distillation from a teacher model with strong cross-domain generalization. The student model is pretrained on a multi-domain corpus comprising speech, environmental sound events, and bioacoustic recordings to further incentivize domain generalization. Furthermore, the central contribution of this work is the integration of modulation-based acoustic features, fused at each Transformer layer through so-called FiLM conditioning~\cite{perez2018film}, which enhances feature disentanglement and representation quality even in low-capacity regimes. Experimental results across multiple bioacoustic benchmarks show that BioME achieves state-of-the-art performance relative to larger models and, in some cases, matches or surpasses its teacher. Ablation studies further validate the contribution of each architectural and training component.

The remainder of this paper is structured as follows. Section 2 reviews related work and background in bioacoustic representation learning. Section 3 introduces the proposed BioME and its components. Section 4 describes the experimental setup, including datasets, training procedures, and evaluation metrics. Section 5 reports and analyzes the results, compares our method with models with varying computational budgets, and discusses its limitations. Finally, Section 6 concludes the paper and outlines directions for future research.

\section{Background and Motivation}

\subsection{Self-Supervised Audio Representation Learning}

Learning meaningful representations from high-dimensional data is a central problem in machine learning. Self-supervised learning (SSL) addresses this by leveraging large collections of unlabeled data and extracting supervisory signals from the data itself. In speech and audio, the dominant SSL approach is masked language modeling, where the model predicts masked portions of the signal using regression, classification, or contrastive objectives.

BEATs~\cite{pmlr-v202-chen23ag}, and its open-source variant OpenBEATs~\cite{bharadwaj2025openbeats}, are patch-based self-supervised audio encoders. They operate by first converting the waveform into a mel-spectrogram, which is then divided into fixed-size time-frequency patches and processed by a Transformer encoder. During pretraining, a subset of patches is masked, and the model is trained to predict the corresponding discrete semantic labels using a classification loss. The training pipeline alternates between refining the encoder and the acoustic tokenizer that produces these labels. Empirically, BEATs and OpenBEATs have shown strong cross-domain generalization and have been applied to several downstream tasks, including bioacoustic benchmarks.

More specifically to the bioacoustic domain, AVES~\cite{hagiwara2023aves} is a self-supervised model adapted from the HuBERT~\cite{10.1109/TASLP.2021.3122291} speech encoder. Unlike patch-based approaches, AVES is a frame-based model that operates directly on the waveform. The audio signal is first processed by a convolutional front end that extracts features at a 50 Hz frame rate, after which a subset of frames is masked and reconstructed by a Transformer encoder trained to predict the missing segments. AVES is pretrained on a mixture of sound events and bioacoustic recordings and has demonstrated strong performance in ecological tasks.

Altogether, both BEATs and AVES contain roughly 90M parameters, which can make them hard to deploy or further improve on resource-constrained devices. Although smaller self-supervised audio models exist (e.g., BYOLA~\cite{niizumi2023byol-a} and SS-AST~\cite{gong2022ssast} with 5M and 23M parameters, respectively), their performance lags behind that of larger encoders, suggesting that model capacity remains a key factor in extracting strong representations. This gap indicates a need for architectures that retain the benefits of large-scale SSL pretraining while remaining lightweight enough for real-world bioacoustic monitoring.

\subsection{Knowledge Distillation}

Knowledge distillation (KD)~\cite{hinton_kd} aims to transfer the behavior of a high-capacity model to a smaller one, enabling the student model to approximate the teacher's generalization ability. In the speech domain, RobustDistiller~\cite{10095480} has demonstrated that knowledge distillation can be used not only to compress a model but also to induce new capabilities, such as increased robustness to noise and reverberation, by introducing strong inductive biases during training via multi-task learning. In addition, USAD~\cite{chang2025usad}, a distilled general-purpose audio encoder, has demonstrated the effectiveness of layer-to-layer distillation, where the student and teacher share the same depth and intermediate representations are directly matched across selected layers. In this work, we adapt both layer-to-layer distillation and DSP-inspired inductive biases in BioME.

\subsection{Bioacoustic Downstream Tasks}\label{sec:background_benchmarks}

Once pretrained, the effectiveness of an audio encoder is typically evaluated through downstream benchmarks that fine-tune the model on supervised tasks under standardized evaluation protocols. In bioacoustics, BEANS~\cite{hagiwara2023beans} is one of the widely used benchmarks, comprising 10 bioacoustic tasks and two general audio tasks (sound event classification and keyword spotting). The bioacoustic tasks span marine mammal, bat, bird, mosquito, and dog vocalization classification, as well as detection tasks for birds, mammals, whales, frogs, and Hainan gibbons. 

Another relevant evaluation setting is beehive acoustic monitoring, where bioacoustic encoders are used to assess honeybee colony health from hive recordings~\cite{11126110}. Honeybees are among the most critical pollinators in global ecosystems and contribute to an estimated 30\% of worldwide food production, making reliable monitoring essential not only for ecological sustainability but also for economic and food security~\cite{potts2010global}. Four common downstream tasks include queen bee presence, colony strength estimation, buzzing classification, and acoustic activity detection. 

\subsection{Modulation Spectrum Signal Representation}

\begin{figure*}[t]
        \centering
        \includegraphics[width=\linewidth]{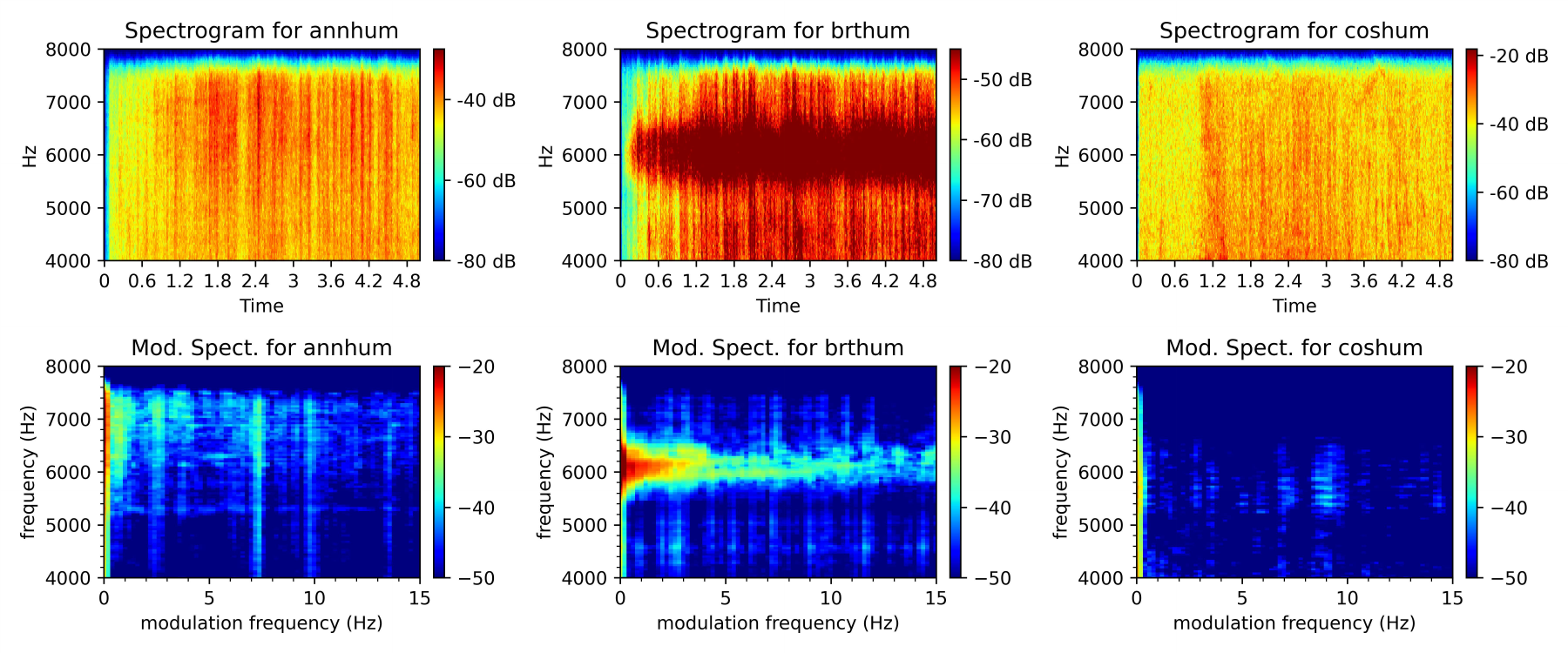}
        \caption{Spectrogram (top row) and the modulation spectrum (bottom row) plots, averaged across samples, for three hummingbird species: Anna’s hummingbird (\emph{annhum}), Broad-tailed hummingbird (\emph{brthum}), and Costa’s hummingbird (\emph{coshum}).}
        \label{fig:spec_entangled_humbd}
\end{figure*}

The modulation spectrum is a signal processing transform designed to quantify the rate-of-change of spectral components~\cite{tiwari2022modulation}. The pipeline for computing the modulation spectrum comprises two main steps: first, the time-domain signal $x(t)$ is mapped to a spectrotemporal representation $X(t, f)$ (e.g., via a short-time Fourier transform). In the second step, a Fast Fourier Transform (FFT) is applied along the time dimension of each spectrotemporal magnitude $|X(t,f)|$ (i.e., the FFT is applied for each frequency bin $f$), yielding the 2-D modulation spectrum representation $X(f_{mod}, f)$. This domain characterizes how specific acoustic frequencies $f$ change at modulation frequencies $f_{mod}$.

This frequency-modulation $\times$ frequency domain is particularly interesting for bioacoustic data collected ``in-the-wild.'' While environmental noise and artifacts often overlap with bioacoustic signals in both time and acoustic frequency, their rates of change typically differ, facilitating their disentanglement in the modulation domain~\cite{falk2009modulation}. This separability enables robust feature extraction even under highly degraded conditions. Furthermore, the representation captures higher-order periodicities, making it highly effective for characterizing animal sounds, which often present strong periodic structures~\cite{tokuda2002nonlinear, burnham2023animal}, such as bird or whale calls. Lastly, the use of modulation-spectral features for acoustic beehive monitoring has been shown to improve performance for colony strength assessment~\cite{11222882}.

To illustrate the advantage of modulation spectral features, we consider the example shown in Figure~\ref{fig:spec_entangled_humbd}. We compute both the spectrogram and the modulation spectrum, averaged across samples, for three hummingbird species with distinct acoustic characteristics: Anna’s hummingbird (\emph{annhum}), Broad-tailed hummingbird (\emph{brthum}), and Costa’s hummingbird (\emph{coshum}). From the spectrograms (top row), we observe a strong structural similarity across the species. Therefore, this can make it challenging for classification models relying solely on spectral information to reliably discriminate between them. In contrast, the modulation spectrum (bottom row) better exposes differences by inserting the modulation-frequency dimension. Our hypothesis is that this representation helps disentangle otherwise overlapping spectral structures, thereby providing more discriminative cues and facilitating species separation.

\begin{figure}[t]
        \centering
        \includegraphics[width=\linewidth]{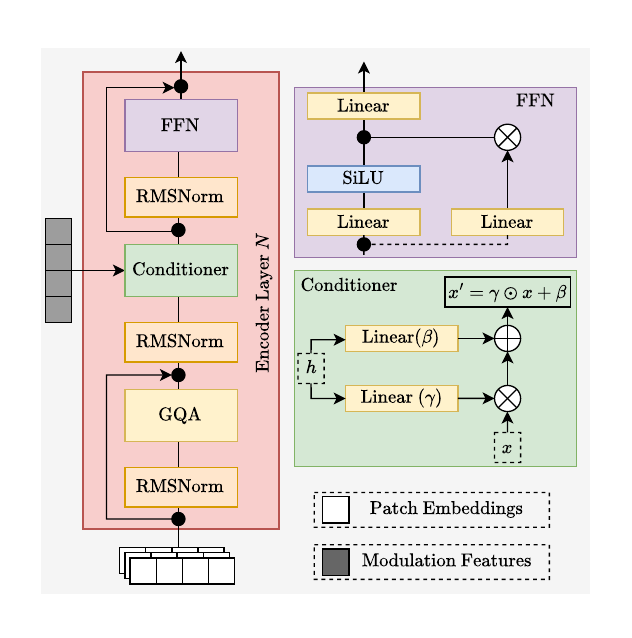}
        \caption{Block diagram of a single Transformer layer in the proposed BioME encoder. Patch embeddings are processed alongside side-channel context features, which are integrated at each layer through the Conditioner module implementing FiLM-based conditioning.}
        \label{fig:biome_components}
\end{figure}

\section{Proposed Method: The BioME audio encoder}\label{sec:benchmark}

In this section, we describe the main components of the BioME audio encoder. Here, we opted for patch-based features, as in the BEATs encoder. First, given a batch of audio waveforms, we compute mel-spectrograms. Next, the spectrogram is split into non-overlapping patches of size $16 \times16$ and linearly projected. The patch embeddings are then fed to the Transformer encoder layers, which we detail next.

\subsection{Transformer Encoder}

Commonly, general-purpose audio encoders use a ViT backbone to process patch embeddings. Traditionally, these networks are equipped with convolution-based relative position embeddings. Given recent advances in the natural language processing (NLP) community, especially in Large Language Models (LLMs), we investigate the use of new components for speech-related tasks. In particular, we adopt some architectural choices from Llama 3.2~\cite{dubey2024llama} for our encoder, as illustrated in Fig.~\ref{fig:biome_components}, and described below.

\subsubsection{Grouped Query Attention (GQA)} GQA is an attention formulation in which multiple query heads share a smaller set of key-value projections~\cite{ainslie2023gqa}. Originally, it was introduced to reduce memory consumption and accelerate decoding in large language models. However, GQA can also provide practical benefits for compact encoders. By tying key-value representations across groups of heads, GQA reduces the number of projection parameters. Empirically, we find that incorporating GQA into our architecture can benefit edge use cases by lowering VRAM usage for attention computation, reducing latency, and improving training stability in low-capacity regimes.

\subsubsection{Rotary Position Embedding (RoPE)} RoPE~\cite{su2024roformer} is employed to encode positional information directly within the attention mechanism. Unlike the learnable positional embeddings commonly used in ViT-based audio encoders, RoPE requires no additional parameters and integrates seamlessly with multi-head attention, enabling more efficient and scalable sequence modeling while reducing computational overhead.

Furthermore, as depicted in Fig.~\ref{fig:biome_components}, BioME adopts additional architectural components inspired by the LLaMA family of models, including the SiLU activation function and RMSNorm layers. We refer interested readers to the technical report~\cite{dubey2024llama} for further details. Lastly, we turn to the integration of modulation-based spectral features via layer-wise conditioning, discussed in the following subsection.

\begin{figure}[t]
        \centering
        \includegraphics[width=0.9\linewidth]{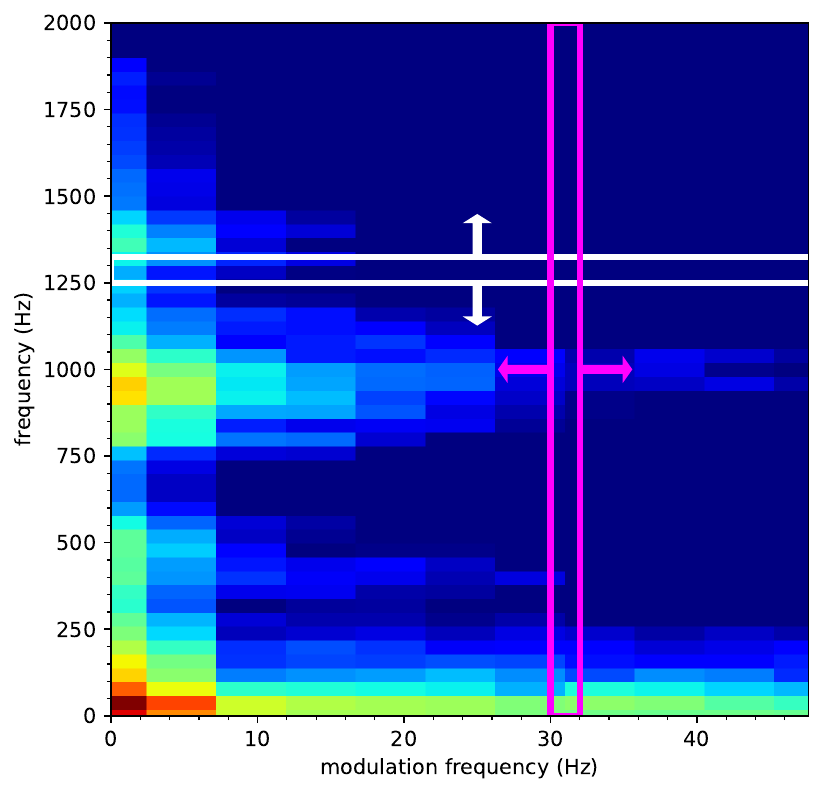}
        \caption{Modulation Spectrogram Average Bands (MSAB) computation. For example, a $256 \times 256$ modulation spectrogram produces a $512$-dimensional MSAB feature vector.}
        \label{fig:msab_computation}
\end{figure}

\subsection{Modulation-features via FiLM}

As previously discussed, the modulation spectrogram is a DSP-based tool to capture the dynamic characteristics of frequency components over time. Its output is a two-dimensional modulation spectrogram with axes corresponding to modulation frequency (x-axis) and acoustic frequency (y-axis). Following prior work~\cite{11222882}, we compute Modulation Spectrogram Average Bands (MSAB) as a compact summary of this representation. Specifically, we average the modulation spectrogram along both axes, as depicted in Fig.~\ref{fig:msab_computation}, and concatenate the resulting vectors that capture band-averaged modulation energies.

The MSAB features are injected into each Transformer layer via a feature-wise linear modulation (FiLM)~\cite{perez2018film} layer, illustrated in Fig.~\ref{fig:biome_components} as the `Conditioner'. Given the (MSAB) context vector $h$, FiLM computes scale ($\gamma$) and shift ($\beta$) parameters, which are applied to the contextualized patch embeddings $x$ as $x' = \gamma \odot x + \beta$. Here, $\gamma$ adjusts the gain of each feature channel, while $\beta$ applies a bias shift. Unlike cross-attention, FiLM provides a lightweight mechanism for attending intermediate representations with side-channel information. Applied post-attention, it acts as a top-down modulation mechanism~\cite{gazzaley2012top}, effectively incorporating DSP-based inductive biases to enable the model to focus on relevant stimuli and ignore distractions based on modulation features.

\begin{table}[htpb]
    \centering
    \caption{Training datasets basic information and data configurations used for our distillation experiments with BioME.}
    \resizebox{0.85\linewidth}{!}{%
        \begin{tabular}{cccccc}
            \toprule
            & & & \multicolumn{2}{c}{\textbf{Data Config.}} \\
            \cmidrule{4-5}
            Dataset & Hours & \#~Utt. & Core & Bio \\
            \midrule
            AudioSet & 4,620.0 & 1,738,788 & \checkmark & \checkmark \\
            FSD50K & 80.7 & 40,966  & \checkmark & \checkmark \\
            VGGSound & 506.2 & 182,615 & \checkmark & \checkmark \\
            iNatSounds & 765.7 & 137,012  & & \checkmark \\

            \midrule

            \textit{Total (Hours)} &  &  & 5,206.9 & 5,972.7 \\

            \bottomrule
        \end{tabular} \label{tab:data_config}
    }
\end{table}

\vspace{-4mm}
\subsection{Layer-wise Knowledge Distillation}

The final component enabling effective training of BioME is knowledge distillation. We adopt BEATs as the teacher model due to its strong performance on both general-purpose audio tasks and bioacoustic benchmarks. To facilitate layer-wise alignment, our encoder mirrors the teacher’s depth (12 Transformer layers) while exploring different configurations to create a thinner architecture. We perform intermediate-layer distillation at the set of layers $K = \{3, 6, 9, 12\}$ for both the teacher and the student, similarly to~\cite{chang2025usad}. When the hidden dimensions differ, a linear projection head maps student representations to the teacher’s dimensionality.

Our distillation objective is inspired by USAD~\cite{chang2025usad} and DistilHuBERT~\cite{chang2022distilhubert}, combining an L1 loss with a cosine similarity term to encourage the student to match both the magnitude and direction of the teacher’s representations. Let $\hat{\mathbf{z}}^{(t)}_{k} \in \mathbb{R}^D$ denote the projected student representation, and $\mathbf{z}^{(t)}_{k} \in \mathbb{R}^D$ the corresponding teacher representation at layer $T$ and patch $K$. The loss is then defined as

\begin{equation*}
    \mathcal{L} = \frac{1}{KT}\sum^{K}_{k=1}\sum^{T}_{t=1} \frac{1}{D}\left\lVert \hat{\mathbf{z}}^{(t)}_{k} - \mathbf{z}^{(t)}_{k}\right\rVert_1 - \log \sigma\left[\cos\left(\hat{\mathbf{z}}^{(t)}_{k}, \mathbf{z}^{(t)}_{k}\right)\right],
\end{equation*}
where $\sigma(\cdot)$ is the sigmoid function and $\cos(\cdot, \cdot)$ is the cosine similarity between two tensors.

\section{Experimental Setup}

\begin{figure*}[t]
        \centering
        \includegraphics[width=\linewidth]{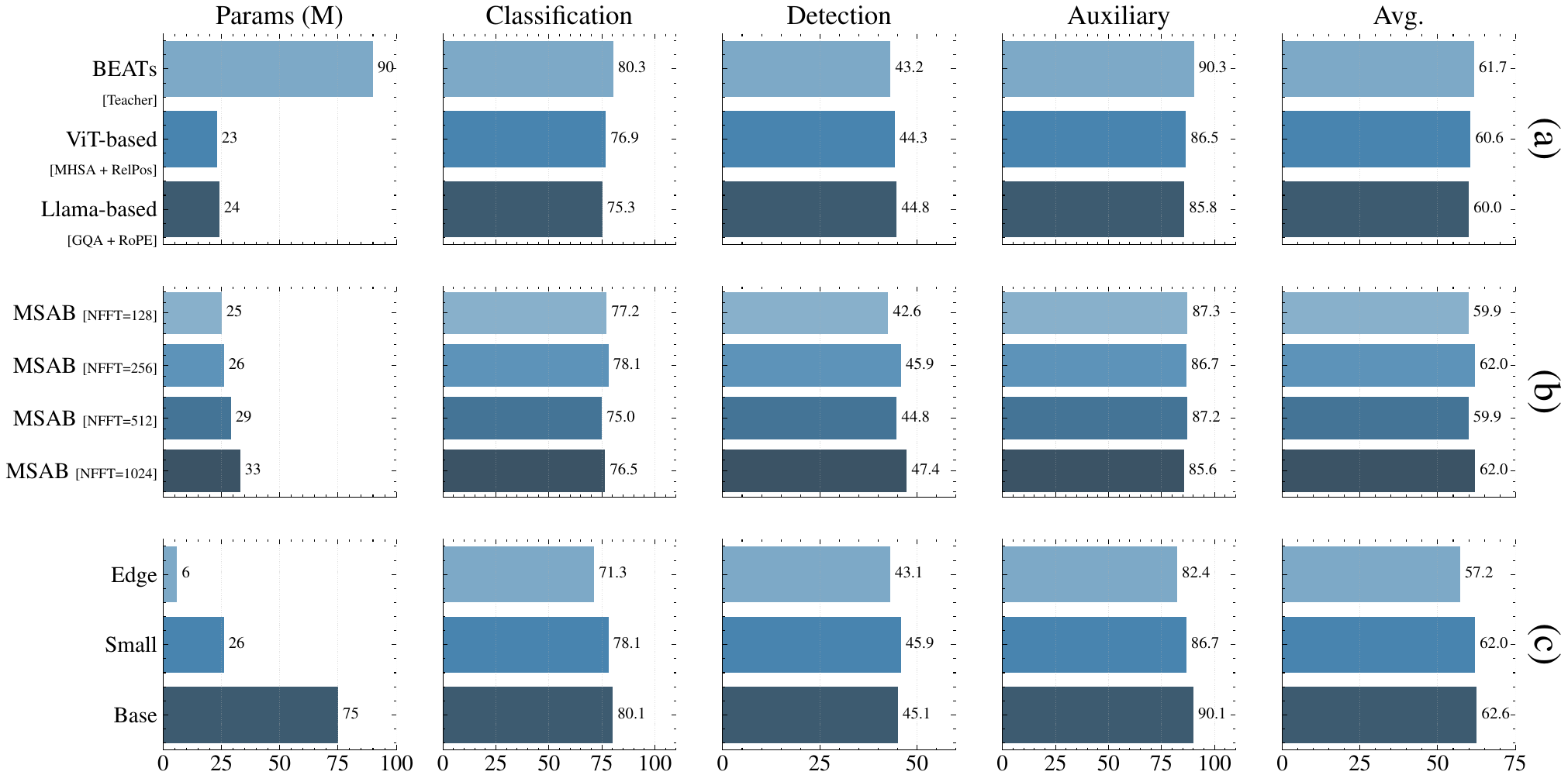}
        \caption{Ablation studies for the proposed BioME. We analyze (a) the student architecture selection, (b) the effect of spectral resolution (NFFT size) on the MSAB features, and (c) model scalability (Edge vs. Small vs. Base). Results are averaged across the classification, detection, and auxiliary tasks of the BEANS benchmark. The `Avg.' column is the average across all bioacoustic tasks (excluding auxiliary tasks).}
        \label{fig:ablation_biome}
\end{figure*}

\subsection{Datasets}~\label{sec:datasets}

For training, we employ four datasets under two configurations, summarized in Table~\ref{tab:data_config}. The first configuration, \emph{Core}, follows a similar dataset composition used in the AVES encoder~\cite{hagiwara2023aves} and includes AudioSet~\cite{7952261}, FSD50K~\cite{9645159}, and VGGSound~\cite{9053174}. The second configuration, \emph{Bio}, extends \emph{Core} by incorporating the recently introduced iNatSounds dataset~\cite{chasmai2024the}, a large-scale bioacoustic collection covering more than 5{,}500 species. However, approximately 70\% of recordings in this dataset are from bird vocalizations. All audio is resampled to \SI{16}{kHz} for consistency with other encoders.

To assess the quality of the learned representations, we utilize several test datasets. Within the BEANS benchmark,datasets representing 12 different tasks are present. These include classification of different marine mammals (\textit{wtkn}), fruit bats (\textit{bat}), birds (\textit{cbi}), mosquitoes (\textit{hbdb}), and domestic dogs (\textit{dogs}). For detection tasks, the datasets contain onset/offset times and cover mammal/bird mixtures (\textit{dcase}), bird dawn choruses (\textit{enab}), Minke whale vocalizations (\textit{hiceas}), frogs and birds from the rainforest fauna (\textit{rfcx}), and gibbon calls (\textit{gib}). Lastly, two auxiliary audio tasks are included and entail sound event classification and keyword spotting. Following~\cite{hagiwara2023beans}, we report accuracy for the classification tasks and mean average precision (mAP) for detection tasks.

For acoustic beehive monitoring, the four tasks from~\cite{11126110} are used, namely: Beehive State (BSTS), Beehive Strength (BSTR), Buzzing Identification (BID), and Voice Activity Detection (VAD). BSTS is a binary classification task designed to detect the presence or absence of a queen bee in a given audio sample. BSTR, a regression task, estimates the colony size; specifically, it predicts the number of hive frames (ranging from 0 to 30) that exhibit at least 80\% honey bee coverage. BID is a multi-class classification task that distinguishes between three acoustic events: bee buzzing, cricket chirping, and ambient natural noise. Finally, VAD is a binary task detecting the presence of human speech within the hive recordings. Following~\cite{11126110}, we use ROC-AUC for the BSTS task, mean absolute error (MAE) for the BSTR task, accuracy for BID, and F1-score for the VAD task as figures-of-merit. In order to rank the methods, the overall score proposed in~\cite{11126110} is also used, where it takes both task performance and model size into account, explicitly rewarding computational efficiency. The score ranges from 0 to 1000, where higher values represent models with the best performance-computational efficiency trade-offs.

\subsection{Hyperparameters and Hardware} We evaluate the scalability of our proposed architecture across three model sizes: an \emph{Edge} variant (6M parameters), a \emph{Small} variant (26M parameters), and a \emph{Base} variant (76M parameters). For the distillation regime, we use the AdamW optimizer with $\beta_1=0.9$, $\beta_2=0.98$, and a weight decay of $0.01$. The learning rate follows a cosine annealing schedule with a linear warm-up: it increases from $10^{-5}$ to a peak of $10^{-4}$ over the first 25,000 steps, then decays to zero over the subsequent 75,000 steps (for a total of 100,000 training steps). We use a global batch size of 128 samples. All models are trained on a single Nvidia H100 GPU. At inference time, however, more modest hardware is sufficient. For example, the peak memory to infer 5 minutes of audio (16 kHz sampling rate, FP-32) for the \emph{Edge}, \emph{Small}, and \emph{Base} models is $1.7$, $3.6$, and $4.3$~GB, respectively.

\begin{table*}[t]
    \centering
    \caption{Main evaluation results for the BEANS benchmark.}
    \makebox[\textwidth][c]{
            \resizebox{\linewidth}{!}{%
            \begin{tabular}{lccccccccccccccc}
                \toprule
                \multicolumn{3}{c}{Upstream Information} & \multirow{2}{*}{Bio} & \multicolumn{5}{c}{Classification} & \multicolumn{5}{c}{Detection} & \multirow{2}{*}{Avg.} \\

                \cmidrule(lr){1-3}
                \cmidrule(lr){5-9}
                \cmidrule(lr){10-14}

                Method & \#P (M) & MMACs/s & & wtkn & bat & cbi & hbdb & dogs & dcase & enab & hiceas & rfcx & gib & \\
                \midrule
                \multicolumn{16}{c}{Supervised Baselines} \\
                \midrule

                rn18 & 12 & 4686 & --- & 75.2 & 44.3 & 35.7 & 69.7 & 66.2 & 16.1 & 32.5 & 28.0 & 6.4 & 16.4 & 39.1 \\ 

                rn50 & 26 & 10616 & --- & 79.9 & 54.8 & 29.5  & 69.6 & 63.3 & 18.3  & 28.2 & 30.4 & 5.5 & 21.5 & 40.1 \\ 

                rn152 & 60 & 29807 & --- & 74.3 & 48.3 & 33.0 & 64.5 & 51.1 & 15.4 & 28.0 & 25.5 & 6.9 & 24.8 & 37.2 \\ 

                vggish & 72 & 864 & --- & 84.7 & 74.3 & 44.0 & 80.8 & 90.6 & 33.5 & 53.5 & 46.3 &  14.0 & 15.0 & 53.7 \\ 
                
                rn152p-s & 60 & 29807 & --- & 83.5 & 60.6 & 58.3 & 70.0 & 79.9 & 30.0 & 52.0 &  32.6  & 9.7 & 30.3 & 50.7 \\ 

                vggish-s & 72 & 864 & --- & 79.4 & 72.3 & 44.5 & 80.1 & 80.6 & 32.5 & 50.9 &  42.8 & 14.3 & 18.6 & 51.6 \\ 

                \midrule
                \multicolumn{16}{c}{Audio SSL Encoders} \\
                \midrule

                BYOL-A~\cite{niizumi2023byol-a} & 5 & 143 & \xmark & 84.4 & 75.0 & 43.3 & 79.4 & 87.1 & 37.2 & 53.4 & 53.8 & 12.6 & 5.5 & 53.2 \\

                SS-AST~\cite{gong2022ssast} & 23 & 10984 & \xmark & 91.2 & 76.6 & 28.5 & 81.2 & 79.1 & 35.2 & 51.0 & 48.8 & 4.5 & 30.7 & 52.7 \\

                USAD~\cite{chang2025usad} & 24 & 1353 & \xmark & 86.1 & 77.6 & 58.3 & 82.0 & 97.1 & 37.2 & 53.8 & 63.1 & 11.3 & 42.0 & 60.9 \\

                Dasheng~\cite{dinkel2024dasheng} & 90 & 2130 & \xmark & 77.3 & 75.5 & 63.1 & 69.9 & 95.7 & 45.6 & 43.9 & 43.8 & 11.7 & 41.8 & 56.8 \\ 

                BEATs~\cite{pmlr-v202-chen23ag} & 90 & 4341 & \xmark & 89.4 & 76.8 & 64.0 & 80.6 & 90.7 & 43.8 & 49.6 & 67.4 & 10.3 & 44.8 & 61.7 \\ 

                OpenBEATs~\cite{openbeats} & 90 & 4337 & \xmark & 87.0 & 76.9 & 64.5 & 84.8 & 95.7 & 46.4 & 12.4 & 42.3 & 68.1 & 49.0 & 62.7 \\ 

                AVES-bio~\cite{hagiwara2023aves} & 89 & 6635 & \checkmark & 87.9 & 74.8 & 59.8 & 81.0 & 95.0 & 39.2 & 55.5 & 62.9 & 13.0 & 28.4 & 59.8 \\ 
                
                GPTM~\cite{10890750} & 89 & 43797 & \checkmark & 91.4 & 77.4 & 54.3 & 81.1 & 94.2 & 45.4 & 62.4 & 65.0 & 12.9 & 34.5 & 61.9 \\ 

                \midrule

                [\textbf{Ours}] BioME (Core) & 6 & 227 & \checkmark & 87.0 & 70.4 & 36.5 & 81.2 & 86.3 & 41.2 & 53.4 & 60.1 & 8.0 & 45.7 & 57.0 \\

                [\textbf{Ours}] BioME (Core) & 26 & 1154 & \checkmark & 89.7 & 71.6 & 54.2 & 81.9 & 87.1 & 38.6 & 59.3 & 69.9 & 11.1 & 50.7 & 61.4 \\

                [\textbf{Ours}] BioME (Core) & 76 & 3427 & \checkmark & 90.9 & 73.2 & 60.9 & 81.6 & 90.6 & 39.1 & 63.3 & 72.0 & 7.1 & 40.6 & 61.9 \\

                \\[-1.2ex] \cdashline{1-16} \\[-1.2ex] 

                [\textbf{Ours}] BioME (Bio) & 6 & 227 & \checkmark & 84.7 & 68.8 & 37.2 & 80.5 & 85.6 & 40.0 & 52.2 & 66.7 & 7.9 & 48.7 & 57.2 \\

                [\textbf{Ours}] BioME (Bio) & 26 & 1154 & \checkmark & 87.0 & 72.2 & 55.3 & 81.2 & 95.0 & 43.0 & 62.1 & 66.2 & 11.1 & 47.3 & 62.0 \\

                [\textbf{Ours}] BioME (Bio) & 76 & 3427 & \checkmark & 91.2 & 74.3 & 59.0 & 81.0 & 95.0 & 42.1 & 62.3 & 69.3 & 10.8 & 40.9 & 62.6 \\

                \bottomrule
            \end{tabular} \label{tab:beans_bench_results}
        }
    }%
\end{table*}

\subsection{Ablation studies}

First, we investigate the effects of our proposed modifications, as summarized in Figure~\ref{fig:ablation_biome}. The ablation study is divided into three experimental sets: (a) Student Architecture, (b) Spectral Resolution, and (c) Model Scaling. For each configuration, we report five metrics: total model parameters, classification accuracy and detection performance (each averaged across five BEANS tasks), auxiliary task performance, and the overall average across all 10 bioacoustic tasks. Except for the Teacher, all models here are trained using the \emph{Bio} partition dataset.

Figure~\ref{fig:ablation_biome}~(a) compares the Teacher model against two student variants: a ``ViT-based'' student (mirroring the teacher's MHSA + RelPos components but with reduced parameters) and our proposed ``Llama-based'' architecture. As observed, both student architectures achieve strong overall performance, remaining competitive with the Teacher despite being approximately four times smaller. We selected the Llama-based architecture for subsequent experiments for two primary reasons: first, its modularity facilitates the integration and testing of side-channel information; second, it offers superior computational efficiency, requiring approximately 5\% fewer millions of multiply-accumulate operations per second (MMACs/s).

Next, we investigate the impact of modulation spectral features and frequency resolution (Figure~\ref{fig:ablation_biome}~(b)). Increasing the FFT size reveals a trade-off between classification and detection performance. Larger FFT sizes (e.g., NFFT=512) improved detection tasks, likely by providing the frequency resolution necessary to distinguish events from background noise. However, this comes at the cost of classification accuracy. We found that an intermediate size (NFFT=256) strikes an optimal balance, maintaining high classification accuracy (78.1\%) while supporting robust detection capabilities.

Finally, utilizing the optimal spectral features, we evaluate the impact of model scaling (Figure~\ref{fig:ablation_biome}~(c)) by ablating two more variants: an \emph{Edge} model (6M params) and a \emph{Base} model (75M params). Increasing the model size to the Base configuration yielded the best overall performance, slightly outperforming even the Teacher model (62.6\% vs 61.7\% average). Conversely, while the Edge model shows a drop in classification performance (from 78.1\% to 71.3\%), it remains a strong candidate for on-device deployment. The edge model still retains the teacher's capability while being approximately $12\times$ smaller than the Base model (and $15\times$ smaller than the Teacher), significantly lowering the barrier for edge implementation.

In summary, our ablation studies helped us develop the final architecture and clearly delineate the benefits of each choice. In particular, the three variants shown in Figure~\ref{fig:ablation_biome}~(c) will be leveraged in the following experiments.

\section{Experimental Results and Discussion}

\subsection{BEANS benchmark tasks}

Next, we compare BioME against established baselines on the BEANS benchmark; results are reported in Table~\ref{tab:beans_bench_results}. We report results for our models trained on both the \emph{Core} and \emph{Bio} data partitions across all three model sizes. We consistently observe that training on the domain-specific \emph{Bio} partition yields superior performance compared to the \emph{Core} partition, confirming the value of in-domain pre-training data. However, we observe only a small performance gap between models trained on the different partitions, particularly for the \emph{Edge} variant. Therefore, this suggests that simply increasing the training data size is not the primary driver of performance for lightweight models; rather, the architectural inductive biases and distillation quality are more critical.

\subsubsection{Efficiency and Scalability} 
\begin{figure}
        \centering
        \includegraphics[width=0.9\linewidth]{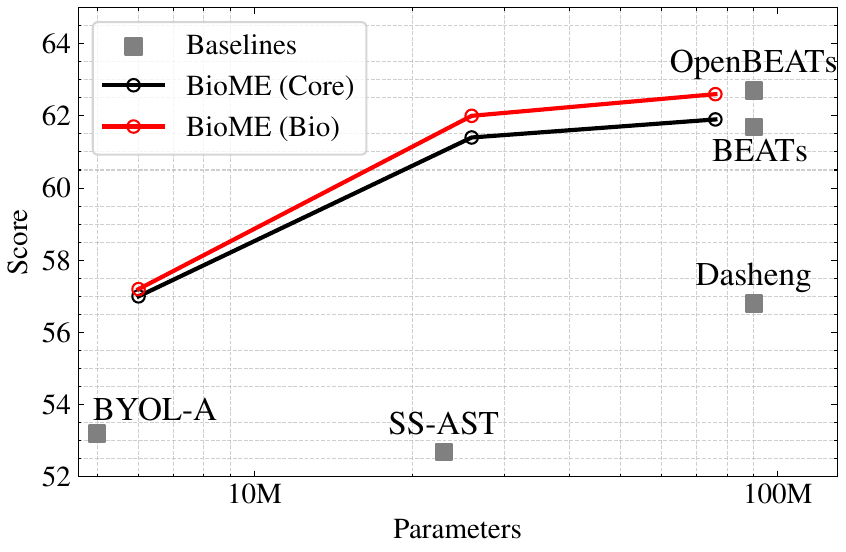}\vspace{-1.5mm}
        \caption{Plot showing the  between parameter efficiency, scaling properties, and training data partition.}
        \label{fig:biome_scale}
\end{figure}

First, we highlight our results regarding model sizes. When comparing models of similar capacity, our distilled architectures significantly outperform standard audio encoders. Specifically, the BioME \emph{Edge} variant (6M parameters) achieves a relative improvement of 7.5\% over BYOL-A (5M parameters). Similarly, the BioME \emph{Small} variant (26M parameters) surpasses SS-AST (23M parameters) with a substantial gain of 17.6\% in overall average score. Figure~\ref{fig:biome_scale} depicts the relationship between model size and performance.

Furthermore, our efficiency gains extend beyond direct size comparisons. The \emph{Edge} model (6M) outperforms significantly larger encoders, including SS-AST ($3.8\times$ larger) and Dasheng ($15\times$ larger, 90M parameters), underscoring the effectiveness of the proposed approach for resource-constrained scenarios.

\begin{figure*}
        \centering
        \includegraphics[width=\linewidth]{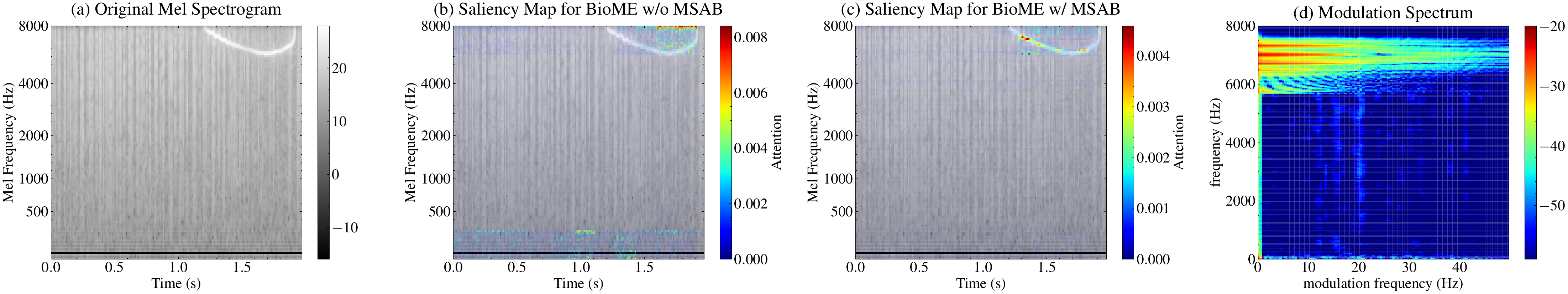}
        \caption{Visualization of a Common Dolphin sample (75003041.wav) from the Watkins dataset. The panels display, from left to right: (a) the input mel spectrogram; (b) saliency maps from the ablated BioME model (without modulation features); (c) saliency maps from the full BioME model; and (d) the corresponding modulation spectrogram used as side-channel information (prior to MSAB transformation).}
        \label{fig:saliency_map}
\end{figure*}

\subsubsection{Teacher-Student Gap} At the \emph{Small} (26M) and \emph{Base} scales (76M), we observe that the student model matches or even slightly exceeds the performance of its teacher, BEATs (62.0\%/62.6\% vs. 61.7\%). While distilled models are typically bounded by their teacher's performance, we hypothesize that the inclusion of side-channel information (MSAB features) introduces critical inductive biases that bridge this gap. This phenomenon, of student models outperforming teachers by leveraging stronger inductive biases or complementary features, has been similarly observed in prior work~\cite{10095480, zhang2025can}. We posit that our proposed recipe, by incorporating these explicit frequency-frequency features, enables the student to disentangle bioacoustic representations more effectively than the teacher.

\subsubsection{Visualizing the impact of MSAB features}

\begin{figure}
        \centering
        \includegraphics[width=0.7\linewidth]{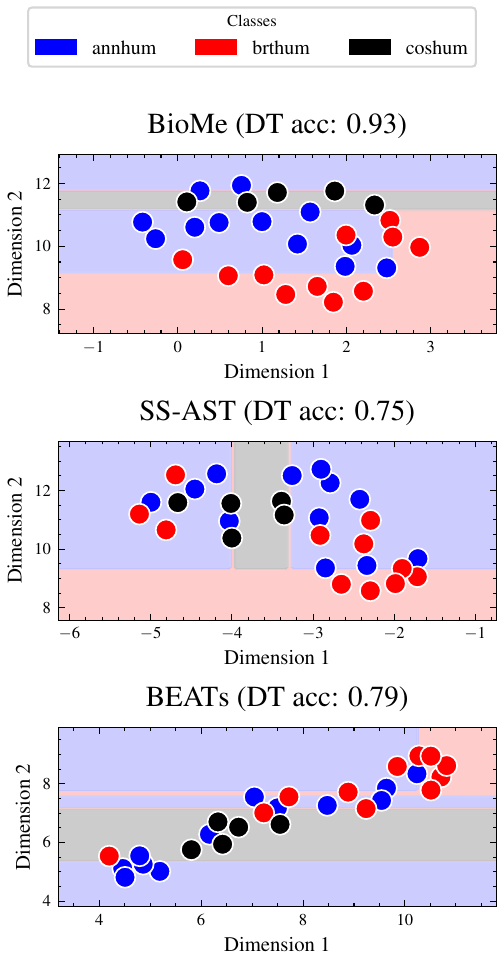}
        \caption{t-SNE visualizations of the latent spaces for BioME (Bio partition, 26M), SS-AST, and BEATs using three hummingbird species (Anna's, Broad-tailed, and Costa's) from the Cornell Bird Identification dataset. To quantify disentanglement, a Decision Tree (max depth=3) was fitted to the 2D projections; the resulting decision boundaries and accuracy scores illustrate the latent disentanglement of each model's representation.}
        \label{fig:tsne_biome}
\end{figure}

Next, to investigate the impact of modulation spectral features on our model, we compare saliency maps with respect to the Mel-filterbank inputs for two configurations: the baseline BioME model (ablated) and the proposed BioME with MSAB features. Figure~\ref{fig:saliency_map}~(a) depicts a sample of a Common Dolphin (\emph{Delphinus delphis}) from the Watkins dataset, characterized by significant underwater background noise. An acoustic event is present between the 1 and 2-second marks. While both models correctly attend to the temporal region of the event (Figure~\ref{fig:saliency_map}~(b) and (c)), the model utilizing MSAB features (Figure~\ref{fig:saliency_map}~(c)) exhibits a distinctly more fine-grained attention pattern. Specifically, it more accurately delineates the harmonic structure and formants of the call compared to the baseline. This observation supports our hypothesis that MSAB features provide important side-channel information, as shown in Figure~\ref{fig:saliency_map}~(d), guiding the model to effectively suppress low-frequency background noise and focus on salient bioacoustic structures during prediction.

In addition, to characterize the learned representations, we analyzed the latent space using the Cornell Bird Identification (\emph{cbi}) dataset, which is part of the BEANS benchmark. We evaluated pretrained models in a zero-shot setting to test the hypothesis that modulation-spectral features help with feature disentanglement, particularly in lower-capacity models. We return to the example presented in Figure~\ref{fig:spec_entangled_humbd} and extract embeddings for the three hummingbird species. Three models are evaluated: BioME (23M), BEATs, and SS-AST. We extract their representation from the last Transformer layer and project it into a two-dimensional space via t-SNE. To quantify separability within this compressed manifold, we fitted a Decision Tree (DT) with a $\text{max\_depth} = 3$ and measured the classification accuracy. The results are illustrated in Figure~\ref{fig:tsne_biome}.

Qualitatively (visual inspection) and quantitatively (DT accuracy), we observe that BioME can produce disentangled latent spaces, which help explain the methodology's overall good performance across different tasks. In this toy example, BEATs takes the second place, followed by SS-AST. While BEATs achieves higher performance on the full 264-class \emph{cbi} task (Table~\ref{tab:beans_bench_results}), where separation occurs in the high-dimensional latent space, this visualization highlights BioME's capacity to maintain distinct class boundaries in a zero-shot manner.

\subsection{Acoustic Beehive Monitoring Tasks}

\begin{table*}
    \centering
    \caption{Results for acoustic beehive monitoring tasks. \textbf{Bold} represents the best and \underline{underline} the second best average results.}
    \makebox[\textwidth][c]{
            \resizebox{0.8\linewidth}{!}{%
            \begin{tabular}{lcccccc}
                \toprule
                \multicolumn{2}{c}{Upstream Information} & \multicolumn{5}{c}{Downstream Tasks} \\
                \midrule
                \multirow{2}{*}{Method} & \multirow{2}{*}{\#P (M)~$\downarrow$} &  BSTS & BSTR & BID & VAD & \multirow{2}{*}{Score} \\
                & & ROC-AUC $\uparrow$ & MAE $\downarrow$ & ACC $\uparrow$ & F1-Score $\uparrow$ & \\
                \midrule
                \multicolumn{7}{c}{DSP-based Features} \\
                \midrule
    
                FBanks & 0 & \makecell{90.04 ($\pm$ 2.40)} & \makecell{10.84 ($\pm$ 0.29)} & \makecell{85.39 ($\pm$ 9.24)} & \makecell{44.53 ($\pm$ 8.46)} & 530 \\
                Spectrogram & 0 & \makecell{86.75 ($\pm$ 2.88)} & \makecell{10.95 ($\pm$ 0.32)} & \makecell{94.49 ($\pm$ 5.02)} & \makecell{51.10 ($\pm$ 7.30)} & 607 \\
                MFCC & 0 & \makecell{71.42 ($\pm$ 8.94)} & \makecell{7.22 ($\pm$ 3.83)} & \makecell{97.95 ($\pm$ 1.22)} & \makecell{30.28 ($\pm$ 5.90)} & 533 \\

                \midrule
                \multicolumn{7}{c}{General-Purpose Audio Encoders} \\
                \midrule

                BYOL-A & 5 & \makecell{63.81 ($\pm$ 7.61)} & \makecell{\underline{3.14} ($\pm$ 0.21)} & \makecell{98.18 ($\pm$ 0.91)} & \makecell{63.07 ($\pm$ 8.83)} & 733 \\
    
                SS-AST ({\small Tiny}) & 6 & \makecell{90.24 ($\pm$ 2.43)} & \makecell{4.44 ($\pm$ 0.28)} & \makecell{91.66 ($\pm$ 2.42)} & \makecell{65.13 ($\pm$ 1.97)} & 832\\
                SS-AST ({\small Small}) & 23 & \makecell{\underline{92.70} ($\pm$ 2.05)} & \makecell{4.45 ($\pm$ 0.16)} & \makecell{96.61 ($\pm$ 2.03)} & \makecell{72.43 ($\pm$ 0.96)} & \underline{910}\\
                SS-AST ({\small Base}) & 89 & \makecell{\textbf{93.91} ($\pm$ 1.44)} & \makecell{4.11 ($\pm$ 0.22)} & \makecell{74.22 ($\pm$ 5.95)} & \makecell{72.19 ($\pm$ 2.32)} & 706\\
                MSM-MAE & 92 & \makecell{89.33 ($\pm$ 1.81)} & \makecell{4.05 ($\pm$ 0.13)} & \makecell{97.24 ($\pm$ 0.81)} & \makecell{73.27 ($\pm$ 1.16)} & 863\\
                M2D & 85 & \makecell{89.22 ($\pm$ 3.45)} & \makecell{3.89 ($\pm$ 0.11)} & \makecell{98.09 ($\pm$ 0.66)} & \makecell{\textbf{74.49} ($\pm$ 1.37)} & 883\\
                CAV-MAE & 85 & \makecell{83.37 ($\pm$ 7.72)} & \makecell{3.42 ($\pm$ 0.22)} & \makecell{93.60 ($\pm$ 2.50)} & \makecell{69.26 ($\pm$ 3.31)} & 797\\
                BEATs & 90 & \makecell{75.22 ($\pm$ 4.30)} & \makecell{\textbf{3.05} ($\pm$ 0.12)} & \makecell{\underline{99.44} ($\pm$ 0.18)} & \makecell{72.82 ($\pm$ 1.06)} & 811\\

                \midrule
                \multicolumn{7}{c}{Bioacoustic-based Audio Encoders} \\
                \midrule

                AVES & 94 & \makecell{90.48 ($\pm$ 1.80)} & \makecell{3.35 ($\pm$ 0.16)} & \makecell{\textbf{99.46} ($\pm$ 0.08)} & \makecell{\underline{74.42} ($\pm$ 0.95)} & \underline{910}\\
                BirdAVES ({\small Base}) & 94 & \makecell{85.32 ($\pm$ 3.64)} & \makecell{3.39 ($\pm$ 0.17)} & \makecell{98.22 ($\pm$ 0.53)} & \makecell{68.59 ($\pm$ 2.93)} & 838\\
                BirdAVES ({\small Large}) & 315 & \makecell{88.98 ($\pm$ 1.04)} & \makecell{3.56 ($\pm$ 0.10)} & \makecell{99.02 ($\pm$ 0.20)} & \makecell{72.72 ($\pm$ 1.61)} & 743\\

                \\[-1.2ex] \cdashline{1-7} \\[-1.2ex] 

                [\textbf{Ours}] BioME (Bio) & 6 & \makecell{92.36 ($\pm$ 3.03)} & \makecell{4.04 ($\pm$ 0.26)} & \makecell{97.19 ($\pm$ 2.62)} & \makecell{68.76	 ($\pm$ 1.16)} & \textbf{917} \\

                [\textbf{Ours}] BioME (Bio) & 26 & \makecell{79.62 ($\pm$ 5.58)} & \makecell{3.63 ($\pm$ 0.39)} & \makecell{97.96 ($\pm$ 0.38)} & \makecell{68.03  ($\pm$ 3.24)} & 833 \\

                [\textbf{Ours}] BioME (Bio) & 75 & \makecell{70.59 ($\pm$ 2.69)} & \makecell{3.59 ($\pm$ 0.11)} & \makecell{98.51 ($\pm$ 0.42)} & \makecell{73.13 ($\pm$ 1.88)} & 770 \\

                \bottomrule
            \end{tabular} \label{tab:abm_main_results}
        }
    }%
\end{table*}

Next, we evaluate our models on the specific applications of Acoustic Beehive Monitoring (ABM), as detailed in Table~\ref{tab:abm_main_results}.

\subsubsection{Performance at the Edge} Remarkably, our smallest model, BioME (\emph{Edge}), achieves state-of-the-art results with a top aggregate score of 917. It is important to note that this composite metric explicitly rewards efficiency, with model size accounting for 20\% of the final score. However, BioME's success is not driven by size alone; it also outperforms larger baselines like AVES and BEATs on the BSTS task. This task is particularly challenging due to its limited data and strict generalization requirements, as the training and test samples are drawn from different hives. These results highlight the feasibility of deploying specialized, resource-efficient audio encoders directly ``in-the-wild'' for edge-enabled beehive monitoring.

\subsubsection{Inverse Scaling Phenomenon} Contrary to the scaling observations in the BEANS benchmark, Table~\ref{tab:abm_main_results} reveals an inverse scaling trend on the beehive monitoring tasks. The \emph{Edge} variant (6M) outperforms both the \emph{Small} (26M) and \emph{Base} (75M) variants, achieving the highest overall score (917 vs. 833 and 770, respectively). This trend is primarily driven by the \textbf{BSTS} task, where the Edge model significantly outperforms its larger counterparts. For the remaining tasks (BSTR, BID, VAD), standard scaling behavior generally applies. We hypothesize that the larger embedding dimension of the \emph{Base} model ($d=768$) compared to the \emph{Edge} model ($d=192$) results in large probing heads that are prone to overfitting in the binary BSTS classification task. Conversely, the compact feature space of the \emph{Edge} model acts as a regularizer, forcing the learning of more robust, generalized representations that are particularly effective for this specific classification task.

\section{Conclusions}

In this work, we introduced \textbf{BioME}, a family of resource-efficient audio encoders designed to enable passive acoustic monitoring on edge devices. By distilling knowledge from the BEATs teacher into a compact student, coupled with modulation-spectral features (via FiLM), we demonstrated that strong inductive biases can effectively compensate for reduced model capacity. Our evaluation on the BEANS benchmark confirms this efficiency: our 6M-parameter \emph{Edge} variant outperforms baselines up to $15\times$ its size, while our \emph{Small} (26M) and \emph{Base} (76M) variants surpass the teacher itself. These results suggest that integrating DSP-inspired features allows student models to better disentangle complex bioacoustic features than purely data-driven methods. In addition, we demonstrate BioME's practical utility for acoustic beehive monitoring tasks, leading to significant performance gains for small models and achieving state-of-the-art performance, highlighting its versatility for diverse ecological applications. In summary, BioME offers a robust and accessible solution for deploying passive acoustic monitoring systems ``in-the-wild''.

\section*{Acknowledgments}
This work was supported by the National Science and Engineering Research Council of Canada (NSERC) via their Alliance program (ALLRP 548872-19).

\bibliographystyle{IEEEtran}
\bibliography{references}

\vfill

\end{document}